# Enhanced Spin-Orbit Torques and Magnetization Switching through Interface Engineering

Yuanmin Du[*]

Department of Materials Science and Engineering, National Tsing Hua University, Hsinchu 30013, Taiwan

The origin of spin-orbit torques generated from the conversion of charge-to-spin currents is of considerable debate. Solid understanding of the physics behind is key to the development of current and voltage controlled switching dynamics in ultrathin heterostuctures. The field free switching observed recently (Phys. Rev. Lett. 120, 117703 (2018)) in a Pt/W/CoFeB structure has intensified such a debate. Here we derive a formula to evaluate a perpendicular effective field generated when the current flows through the heterostructure, considering the large resistivity difference between the two normal metal layers and the chemical potential gradient created at the interface. Together with recent X-ray photoelectron spectroscopy findings at the interface of a Pd/CoFeB structure, we conclude that a new torque generated may play a key role in the field free switching. The model and mechanism proposed agrees with previous reports on spin current injection and field free switching using different interface engineering methods.

[*] E-mail: ynmin.du@gmail.com

Electrical manipulation of magnetization is becoming increasingly important because it has great potential for memory and logic device applications. In recent years, a method called spin-orbit torque (SOT) induced magnetization switching has attracted great attention [1-3]. Spin Hall effect and Rashba effect are reported to play a key role in the process [4-8]. In a simple normal metal (NM) /ferromagnet (FM) structure, an applied in-plane current in NM can generate a spin current, through which magnetization of the adjacent FM can be switched with the assistance of an external magnetic field. For device applications, a structure with the merits of low switching current density, field-free and perpendicular magnetic anisotropy (PMA) properties is desirable. By using an asymmetrical geometrical shape, a ferroelectric material, or a magnetic exchange bias generated by an antiferromagnetic (AFM) material, a field-free switching can be achieved [9-11]. Most recently, a field free switching has also been realized in a Pt/W/CoFeB structure, which shows great challenge to the conventional SOT-based switching mechanism, as reported by Ma et al. [12]. A magnetic switching has even been observed with both the anti-damping-like torque $\tau_{DL}$ and field-like torque $\tau_{FL}$ at zero, measured in a harmonic measurement. Although a model based on movement of domain walls as a result of the competing current from the two NM metals has been proposed, the detailed physics behind is lacking. In this work, we intend to revisit this issue with a mechanism which has been ignored in the previous reports. We show that spin current injection and magnetization switching in a NM/FM heterostructure can be modulated and enhanced through interface engineering.

When an electron current flows in a metal, its behavior can be described by a simplified Drude model by assuming fixed ions with free electrons [13]. In many conditions, a constant resistivity can be used in calculations of a bulk metal material, and Ohm's law works. When the film thickness goes down to a few nm or even the sub-nm scale, along with material properties like work function and electron effective mass, the resistivity changes dramatically. It is even more different for the case of a double-layer thin film structure. As the film thickness approaches the mean free path for electron-phonon collisions, the preponderant scattering at the boundaries can lead to a significant change of the electron transport behavior. For a heterostucture shown in Fig. 1(a), with Metal 1 having a much higher resistivity than that of Metal 2, i.e., $\rho_1 \gg \rho_2$, reflection and backflow of electrons at the interface occurs. The existence of nano-crystalline or nano-grain structures enhances the ballistic scattering of electrons (at the boundaries/interfaces), and as a result, a circular current can be formed, as shown in Fig. 1(b). Uniformity, surface roughness and discontinuity are three known issues in the fabrication process of thin metal films, which could further enhance such a process. An effective magnetic field is then created by each current loop, which is known as the Ampere's law. As a whole effect, a magnetic field $\mathbf{H}_y^*$ is generated in the -y direction with electron flowing in the x direction. The closed loop is no necessary in a regular circle shape, as shown in Fig. 2, which is determined by the material properties. Electrons are continuously reflected at the boundaries formed in the material. As reported by Demasius et al. [14], the nano-crystalline characteristic of the tungsten film can even be controlled through the incorporation of oxygen, which work shows one way of controlling the metal material property. For the structure Pt(3 nm)/W($t_w = 0.7 - 1.6$ nm)/CoFeB(1 nm) reported by Ma et al. [12], the W layer has a β-W phase which is known to have a much higher resistivity than that of Pt. Simply based on Ohm's law, the ratio of distributed currents in Pt to that in W increases with decreasing the W thickness. After further considering the increasing of W resistivity with decreasing of the thickness [15], the resistivity difference between Pt and W would be even larger. The effective magnetic field $\mathbf{H}_y^*$ generated by the in-plane currents could be very significant in their structure.

Spin currents can be generated by flowing a charge current through the heavy metal. Spin-up and spin-down electrons have two different electrochemical potentials, and carry current independently [16]. The spin-dependent chemical potentials µ↑,↓ as a function of the position for a typical FM/NM interface is shown in Fig. 3(a). A recent report by Du et al. [17] shows that highly pure metallic Pd has been formed in the Pd/CoFeB structure compared to that of the Pd/CoFe structure, leading to a large enhancement of the spin Hall angle (SHA) of Pd. Time-resolved X-ray photoelectron spectroscopy (XPS) spectra measured by synchrotron based light sources using low $Ar^+$ sputtering energy further shows that there are two components of the Pd 3d peak: metallic component (M-Pd) and oxidic component ($PdO_x$, x>0 and is variable) in the Pd layer. The metallic component increases gradually from the interface and saturates in the deep Pd region, and the oxidic component changes in an opposite way, and as a result, a chemical gradient is formed at the interface, as shown in Fig. 3(b). Compared to Fig. 3(a), the chemical potential difference between FM and NM increases from Δµ to Δµ^.

The spin injection can be modulated through change of chemical potential at the interface. On the basis of the work by Valet and Fert [18], we have the spin current density gradient as the following expression:

$$\frac{\partial J_s}{\partial z} = \frac{\mu_\uparrow - \mu_\downarrow}{\lambda_{sf}^2} \frac{\sigma_s}{e} \tag{1}$$

where $\sigma_s$ is spin conductivity, $\lambda_{sf}$ is spin diffusion length, $\mu_\uparrow$ and $\mu_\downarrow$ are spin-dependent chemical potentials. Shown from Fig. 3(a) to Fig. 3(b), a change of the chemical potential profile leads to a change in spin densities as well as the spin current. The spin current density can be enhanced by the chemical gradient, with $J_s \propto \frac{\sigma_s}{e} \frac{\partial \bar{\mu}}{\partial z}$, where $\bar{\mu} = \alpha \mu^\uparrow + (1-\alpha)\mu^\downarrow$, and $\alpha$ is the percentage of spin-up electrons. Compared to that shown in Fig. 3(b), an opposite potential gradient is shown in Fig. 3(c), which can also be realized experimentally.

The gradient of spin densities can induce an extra torque for current-induced magnetization switching. Roles of nonequilibrium conduction electrons on the magnetic dynamics has been discussed by Zhang and Li [19]. Spin-orbit torques generated through the injection of spin electrons from NM to FM are widely reported to be the source for magnetization switching, for which an external magnetic field is generally required. Here we propose a new torque $\tau_n = -c\,\boldsymbol{\sigma} \times \left(\mathbf{H}_y^* \times \frac{\partial \mathbf{J}_s}{\partial z}\right)$ in the NM/FM system, where $\mathbf{H}_y^*$ is the effective magnetic field generated by the circular electron currents described in Figs. 1 and 2, $\boldsymbol{\sigma}$ is the direction of the spin current (simply treated as along y or –y directions), and $c$ is a constant. We will use this torque to discuss field-free switching in a PMA system. The Landau-Lifshitz-Gilbert (LLG) equation can be written as:

$$\frac{\partial \mathbf{m}}{\partial t} = -\gamma\, \mathbf{m} \times \mathbf{H}_{eff} + \alpha\, \mathbf{m} \times \frac{\partial \mathbf{m}}{\partial t} - \gamma\, c\, \mathbf{m} \times \left(\boldsymbol{\sigma} \times \mathbf{H}_y^* \times \frac{\partial \mathbf{J}_s}{\partial z}\right) + \gamma\, \mathfrak{t}_d\, \mathbf{m} \times \left(\boldsymbol{\sigma} \times \frac{\partial \mathbf{J}_s^*}{\partial z}\right) + \gamma\, \mathfrak{t}_f\, \mathbf{m} \times \boldsymbol{\sigma} \qquad (2)$$

where γ is the gyromagnetic ratio, α is the Gilbert damping parameter, and $\mathbf{H}_{eff}$ is the sum of $\mathbf{H}_{ext}$ and the out-of-plane demagnetization field. $\mathfrak{t}_d$ and $\mathfrak{t}_f$ are related to an anti-damping-like term and a field-like term respectively. To differentiate with $\frac{\partial \mathbf{J}_s}{\partial z}$ in NM, we use $\frac{\partial \mathbf{J}_s^*}{\partial z}$ to describe the spin current gradient in the FM. The first two terms are a precession term and a damping term respectively. The third term is related to the new torque introduced here, which is a distanced torque in nature. The fourth term is a spin transfer torque, generated through a direct momentum transfer from the injected spin electrons into the FM, different to the term $\mathbf{m} \times \boldsymbol{\sigma} \times \mathbf{m}$ described by Liu et al. [20,21], which we found great difficulty in explaining the various phenomena observed in different reports. For the effective field $\mathbf{H}_y^*$ in the third term, it can be described as:

$$\mathbf{H}_y^* = \sum \frac{\mu_0 I_i^*}{2 r_i} \qquad (3)$$

where $I_i^*$ is the circular current, $r_i$ is the equivalent radius of the closed circle, and $\mu_0$ is the permeability of free space [22]. In the end of this letter, we will discuss different magnetic dynamic behaviors using the third and fourth terms.

For the Pt/W/CoFeB structure reported by Ma et al. [12], a field free switching can be obtained with a reduced net spin current compared to those with W and Pt alone. This even takes place when the spin current becomes zero, indicating that the injected spin current from NM to FM might not be the source for magnetization switching, i.e., the conventional SOT switching mechanism does not work. As a common impurity element in film fabrication process, a small amount of oxygen can play a key role. Compared to W, Pt has a much higher oxidation resistance, which means that a WOx (x>0, variable) layer could be formed at the NM/FM interface, resulting in a shift of chemical potential in this region, similar to that reported by Du et al. (Pd/CoFeB, [17]), the condition shown in Fig. 3(b). In fact, an enhancement of spin signals due to native oxide formed at Ag/Permalloy interface has also been reported [23]. Using naturally oxidized Cu oxide, a large enhancement of spin-current injection efficiency has been demonstrated by An et al. [24], in which a gradual change of oxygen concentration has been mentioned. Through modulation of oxygen concentration and gradient in a Ta/CoFeB/Ta(O) structure, a field-free switching has been realized by Yu et al. [9].

Although a model based on domain wall movements due to competing spin currents has been proposed by Ma et al. [12], the detailed physic behind is unclear, e.g., shifting of the hysteresis loops with increasing of the charge current, which looks like that an additional perpendicular field is applied and increases with the current. Based on the above analysis, the new torque (the third term in Eq. (2)) introduced here meets all the requirements. The zero field switching takes place at smaller W thicknesses and disappears when further increasing of the thickness to above 1.3 nm, which we believe is the effective distance of the third term torque. Out of this range, an external magnetic field is needed for a switching, i.e., the spin transfer torque described by the fourth term in Eq. 2 will take the role. Due to the great difficulties in resistivity measurements for each layer in a heterostrucure, especially when the thickness goes down to the sub-nm range, the threshold switching current density calculated would have great uncertainty in their report. Discontinuity or islands (and roughness) could be formed in their films when the thickness goes down to below 1 nm. To acquire a good PMA, a high temperature process is always needed in their structures. Inter-diffusion of atoms between different layers could take place, leading to the creation of more scattering centers, which make a quantified calculation even more difficult. Nevertheless, an increase in current density can increase the new torque $\tau_n$ magnitude, and a magnetic switching will occur when its magnitude is large enough to overcome the energy barrier.

In comparison to a NM/FM structure discussed above, an additional capping layer is frequently used by different groups for the purpose of spin current injection enhancement [25,26]. Since the capping layer always suffers from different levels of oxidation, we describe the structure as NM/FM/M(O) here (Fig. 3(d)). Oxidation of the capping layer often takes place from the surface and extends to certain depths. Compared to the surface, a more metallic behavior is shown at the lower regions. The chemical potential gradient (combined with the screening effect) generated at the top layer can result in an adding effect to the existing gradient formed beneath. For a Pt/Co/Ta(O) structure reported by Woo et al. [25], this effect results in a significant enhancement of the SHA, considerably larger than the sum of SHA for Pt and Ta. Further increasing the Ta(O) capping layer thickness can reverse such an effect, which means that the screening effect becomes weaker and weaker when a thick enough metallic layer is formed (and the thickness is increasing) at the lower regions [26]. Controlled oxidation through the capping layer can lead to positive or negative effects to the heterostructure, which is an engineering work.

Now we discuss the chiral switching (clockwise and anticlockwise, in the reports) behavior for different structures based on the new torque. Pt and Ta are two typical metals which have positive and negative SHA values respectively. In the model described in this report, without an external magnetic field, the chirality of field-free switching obeys that described in Fig. 4. Under the same chemical potential gradient condition, the switching sequence follows a clockwise characteristic for the Pt-based structure and anticlockwise characteristic for the Ta-based structure, as shown in Figs. 4(a) and 4(b) respectively. The effective field $\mathbf{H}_y^*$, the direction of the spin moment (along y or –y direction) and the chemical gradient together determine the chirality behavior.

For a NM/FM system with PMA, the model and the mechanism proposed here have various experimental consequences:

- The existence of the new torque will lead to a shift in anomalous Hall effect (AHE) loops when a current is flowing through the heterostructure. An asymmetric rather than a symmetric switching characteristic will be presented. This behavior has been observed in various reports [7,9,12]. As demonstrated by Ma et al. [12], increasing the charge current density increases the shift, and finally leads the loop shifting to one side, and then field free switching takes place.

- The magnitude of the new torque depends on interface conditions. Increasing the resistivity difference and the chemical potential gradient are two ways to achieve a large enough torque to realize an efficient field free switching. Interface engineering using oxygen is one solution [9,14,27]. By incorporation of oxygen into tungsten, not only the resistivity, but also the crystalline property as well as the grain size can be controlled [14]. Change of oxygen profile at the interface with the assistance of boron has recently been demonstrated [17].

- Functionally, the term $\mathbf{H}_y^* \times \frac{\partial J_s}{\partial z}$ can be treated as a magnetic field or bias along the x direction. As reported by Yu et al. [9], a torque created by an in-plane electric field as a result of a non-uniform charge distribution in a wedge-shaped Ta(O) capping layer breaks the so-called symmetry in a Ta/CoFeB/Ta(O) structure, and a field free switching has been realized. In this case, the chemical potential gradient along the z direction has also been changed within the controlled oxidation process. By incorporating ferroelectric PMN-PT into a Pt(4 nm)/Co(0.4 nm)/Ni(0.2 nm)/Co(0.4 nm) structure, a field free switching through a spin current density gradient created along the x direction has also been demonstrated by Cai el al. [11]. For a ferroelectric material, when an electric field is applied laterally in their case, a potential gradient along the z direction has been created simultaneously. Due to the large resistivity difference (and roughness) at the interface, the interface scattering can be very significant. In principle, the field-free switching described in these two works is similar to the situation described in this letter. For the field-free switching by inserting an AFM layer by Fukami et al. [10], it is also assisted by a field or bias along the x direction. Interestingly, a shift of the hysteresis loops with changing the current has also been demonstrated. The new torque may help to understand the AFM behavior presented in such heterostructures.

- Through an oxidation process of the FM (or the interface region), change of the switching chirality has been reported by Qiu et al. [27]. The AHE loops were measured with an in-plane magnetic field applied along the x direction. In this work, the chiral switching behavior can also be changed through controlling the thickness of the $SiO_2$ capping layer of a Pt/CoFeB/MgO/SiO$_2$ structure. Through an oxygen engineering process, the chemical potential gradient across FM has been changed, leading to a sign change of the $\frac{\partial J_s^*}{\partial z}$ described in the fourth term of Eq. (2). The energy band alignment at the interfaces changes in the process [28].

- Different metals have different oxidation resistances. Pt and Pd are two materials with high oxidation resistances [29,30]. Compared to Ni, Fe is more easily to be oxidized [31]. There is also a large difference between Cu, and Al, Ta. The latter two are frequently used as a capping layer in a NM/FM based structure, due to their much lower oxidation resistances. The competition between different metals for limited oxygen can lead to a re-distribution of oxygen at different layers. Through control of the chemical potential gradient, the spin injection efficiency can be enhanced. As reported by Zhang et al. [32], the interface transparency for spin current injection can be largely enhanced by inserting certain materials.

- When the third term is not significant enough, the spin transfer torque described by the fourth term in Eq. (2) will dominate a magnetic switching process with the assistance of an external magnetic field for a PMA system.

- With an electrical current flowing through the system, the coupling of charge, heat and spin currents can give rise to thermoelectric or joule heating effects, such as the anomalous Nernst (ANE) and spin Seebeck (SSE) effects [33,34], which will be discussed and reported separately.

In conclusion, we have shown theoretically that spin injection and magnetic switching in a NM/FM system can be modulated through engineering the interface resistivity and the chemical potential gradient. This work may help understand the spin-orbit torque related phenomena. We also hope that our findings may pave the way towards configurable memory and logic device applications.

The author acknowledges support from National Tsing Hua University, and valuable help by Yufan Li from Johns Hopkins University for providing experimental details in their work.

---

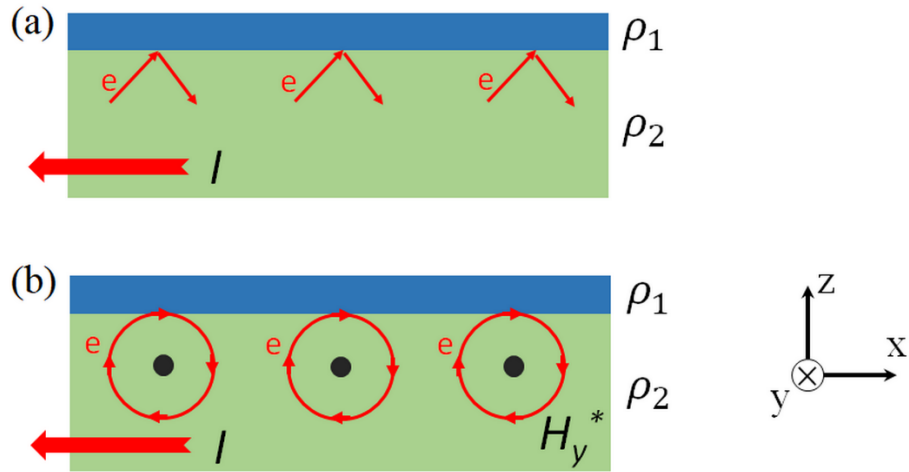

FIG. 1. (a) Electron reflection and scattering at an interface. A hetero-structure is formed. Resistivity $\rho_1 \gg \rho_2$. (b) Circular currents formed due to the electron reflection effect at the boundaries. An out-of-plane effective magnetic field $H_y^*$ is formed based on the right hand rule.

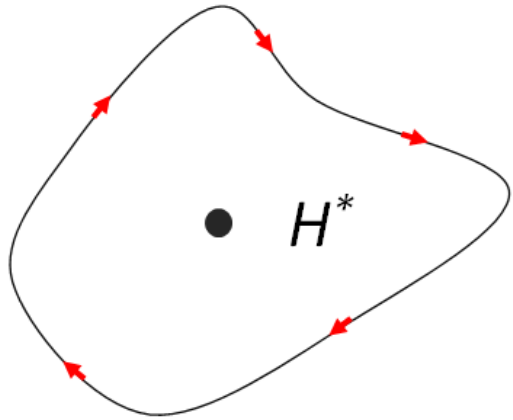

FIG. 2. A non-regular current loop and the effective magnetic field generated. The red arrows depict the electron flow direction.

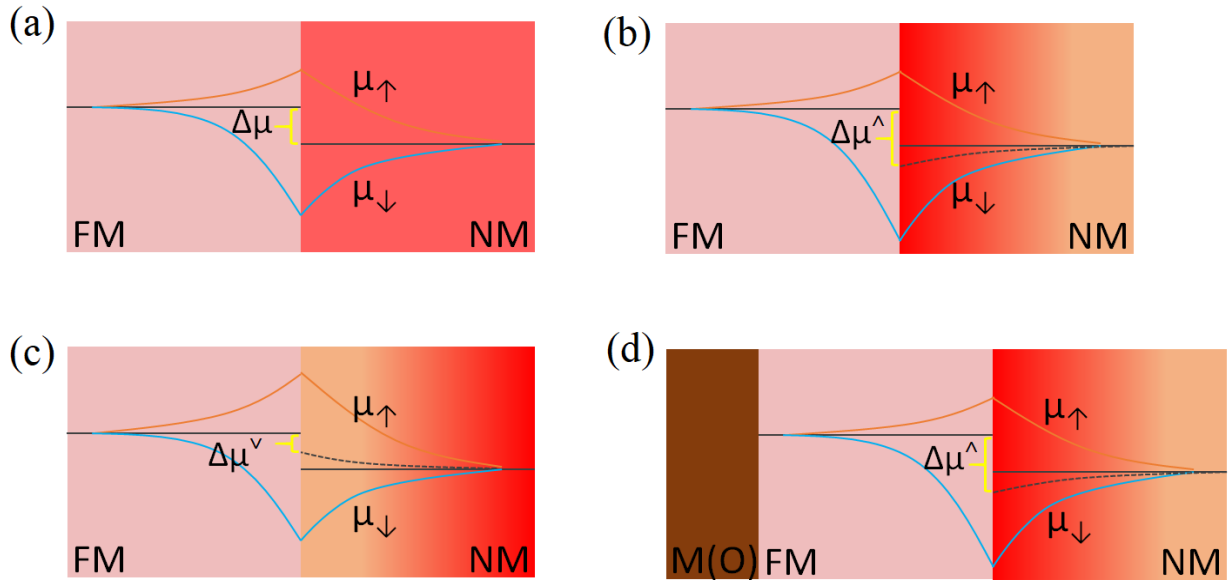

FIG. 3. Schematic of the spin-dependent chemical potentials $\mu_{\uparrow,\downarrow}$ across an FM/NM interface. (a) No potential gradient is formed in the NM, and a chemical potential difference $\Delta\mu$ between FM and NM is formed. (b) A potential gradient is formed in the NM, and an increased potential difference $\Delta\mu^{\wedge}$ is formed. (c) An opposite potential gradient to (b) is formed in the NM, and a decreased potential difference $\Delta\mu^{\vee}$ is formed. (d) The effect of a metal capping layer M(O) to the structure shown in (b). See the text for more details of (d). FM – ferromagnetic metal. NM – normal metal. M(O) – metal incorporated with oxygen.

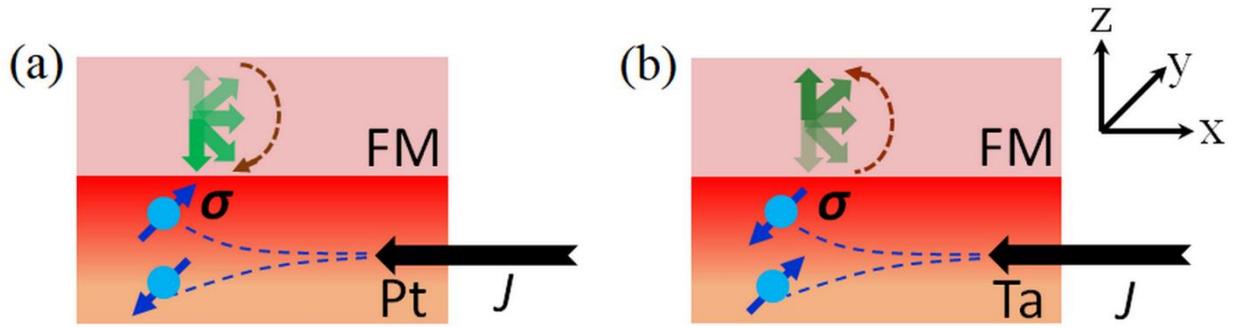

FIG. 4. The different chiral switching characteristics for a field-free switching. (a) Clockwise switching for a Pt based structure. (b) Antilockwise switching for a Ta based structure. An effective magnetic field $H_y^*$ along the -y direction is not shown.